\documentclass[%
 aip,
 amsmath,amssymb,
 reprint,%
]{revtex4-1}

\usepackage{graphicx}
\usepackage{dcolumn}
\usepackage{bm}
\usepackage[colorlinks,linkcolor=blue,anchorcolor=blue,citecolor=blue,urlcolor=blue]{hyperref}
\usepackage[utf8]{inputenc}
\usepackage[T1]{fontenc}
\usepackage{mathptmx}

\begin{document}

\preprint{AIP/123-QED}

\title[Cooling of a levitated nanoparticle with digital parametric feedback]{Cooling of a levitated nanoparticle with digital parametric feedback}

\author{Yu Zheng}
\author{Guang-Can Guo}%
\author{Fang-Wen Sun}
 \email{fwsun@ustc.edu.cn}
\affiliation{CAS Key Laboratory of Quantum Information, University of Science and Technology of China, Hefei 230026, China
}%
\affiliation{Synergetic Innovation Center of Quantum Information and Quantum Physics, University of Science and Technology of China, Hefei 230026, China
}%
\date{\today}

\begin{abstract}
The motion control of a levitated nanoparticle plays a central role in optical levitation for fundamental studies and practical applications. Here, we presented a digital parametric feedback cooling based on switching between two trapping laser intensity levels with square wave modulations. The effects of modulation depth and modulation signal phase on the cooling result were investigated in detail. With such a digital parametric feedback method, the centre-of-mass temperature of all three motional degrees of freedom can be cooled to dozens of milli-Kelvin, which paved the way to fully control the motion of the levitated nanoparticle with a programmable digital process for wild applications.
\end{abstract}

\maketitle
Levitated optomechanics has become a promising platform for ultra-weak force detection\cite{Hempston2017force,geraci2010force,ranjit2015force,ranjit2016force,hebestreit2018freefall}, microscale thermodynamic investigation\cite{gieseler2018MicroTherm,li2010Brown,millen2014nanoscaleTemp,rondin2017Karmers,gieseler2014non-equilibrium} or quantum hybrid systems\cite{neukirch2015NVhybrid,hoang2016NVhybrid,yin2013spinCouple,scala2013matterWave}. Almost all of levitated optomechanics systems need a method to precisely control the motion of the trapped object to fulfill the research and application requirements. A lot of control schemes have been realized, such as cavity-assisted cooling\cite{Windey2019Cavity3D,Deli2019Cavity3D,Kiesel2013Cavity,Millen2015Cavity}, radiation pressure method\cite{Li2013}, electrostatic cooling\cite{Tebbenjohanns2018electric,iwasaki2018electric,conangla2018electric}, and torsion, rotation control\cite{arita2013rotation,Hoang2016torsion,Kuhn2017rotation,Reimann2018rotation,Ahn2018rotation}. Among the control methods, parametric feedback control\cite{Gieseler2012PFC,Vovrosh2017PFC} has become the most popular method. It can be conveniently deployed in laser trapping based levitated system with excellent performance. Therefore, parametric feedback control has been already used in static force detection with free-falling nanoparticle\cite{hebestreit2018freefall}, the verification of recoil heating\cite{jain2016recoil}, the demonstration of phonon laser\cite{Pettit2019phononlaser} and pre-cooling for micro-Kelvin cooling\cite{Tebbenjohanns2018electric}.

In general, the key of the parametric feedback control is to modulate the intensity of trapping laser according to the motions of the levitated particle. In previous schemes, the feedback signal was usually generated by doubling the frequency of the motion signals of the levitated particle with some phase shift.\cite{Gieseler2012PFC} The feedback control loop is usually assembled with analog circuit and the output control signal is a continuous sinusoidal wave, which may limit the application in a complex motion control of the levitated particle. Here, we present a digital parametric feedback method to control the motion of the levitated particle based on a field-programmable gate array (FPGA). It can be plugged into the feedback control loop to deal with some complicated signal processing algorithms.
Based on the digital control circuit, we can apply a square wave signal modulation to precisely control all three motional degrees of freedom of the levitated particle, where the signal phase was modulated to control the heating or cooling processes for the levitated particle. By adjusting the modulation depth, the levitated particle can be cooled to dozens of milli-Kelvin at low pressure. Such a programmable digital parametric feedback method can be further applied in the control of the motions of levitated particle with complex circuits.

\begin{figure}[b]
	\includegraphics[width=0.45\textwidth]{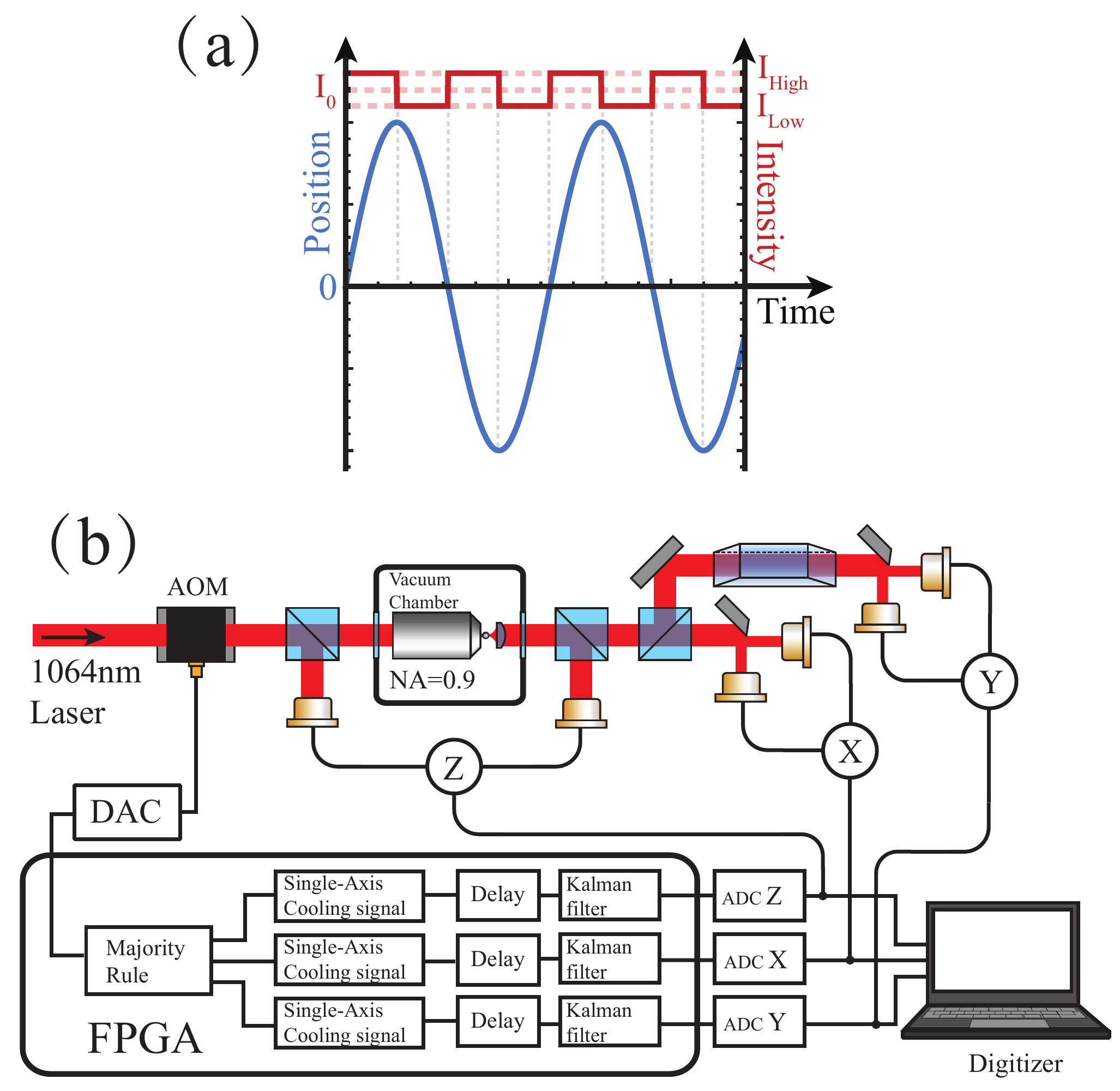}
	\caption{(a) Schematic diagram of single-axial 2-level digital parametric feedback cooling. (b) Experimental setup for tri-axial 2-level digital feedback control. The three-dimensional position signals measured by three sets of balanced photodetectors are sent to the FPGA board to generate the feedback signal and simultaneously recorded by a digitizer. The feedback signal is sent to the AOM to control the intensity of trapping laser. DAC: Digital-to-Analog Converter. ADC: Analog-to-Digital Converter.}
	\label{fig:1}
\end{figure}

Similar to the previous version of analog parametric feedback control, the motions of the particle are controlled by adjusting the intensity of the trapping laser. To cool the centre-of-mass motion (CM) temperature, the laser intensity should be increased (reduced) when the levitated particle is moving away from (towards to) the equilibrium point. But, instead of continuous intensity varying, the laser intensity is only digitally switched between two levels, the high level $I_H$ and the low level $I_L$, with square wave modulations as shown in Fig. \ref{fig:1}(a). For simplicity, to discuss the details of the 2-level digital modulation, we focus on the energy varying of the levitated particle. Under the digital parametric feedback control, when the oscillating particle is moving away from the equilibrium point, the laser intensity is switched to the high level $I_H$. When the particle reaches the peak point, all the motion energy is converted to potential energy which is stored in the light field. At this time point, the laser intensity is reduced to $I_L$ till the oscillating particle returns back to the equilibrium point. The total energy will be lost as the dissipation of potential energy. Here, by setting a modulation depth as $\eta=(I_{H}-I_{L})/I_{0}$ with ${I}_{0}=\left(I_{H}+I_{L}\right) / 2$, we can get the energy reduction during one cycle of the oscillation from the modulation, which is written as
\begin{equation}
\Delta \mathrm{E}=-m \eta \omega_{0}^{2} X_{0}^{2} \text{,}
\end{equation}
where $m$ is the mass of the oscillating levitated particle, and $\omega_0$ is its natural frequency under $I_0$. Moreover, the cooling effect of feedback modulation can be regarded as an additional damping $\delta \Gamma$\cite{Gieseler2012PFC}. Since the total energy loss in each cycle is $\Delta \mathrm{E}_{damp}=-m {\pi}(\Gamma_0+\delta\Gamma) \omega_{0} X_{0}^{2}$ where $\Gamma_0$ is the air damping rate, we can obtain
\begin{equation}
\delta \Gamma=\eta \omega_{0}/{\pi}\text{.}
\label{DGamma}
\end{equation}

Therefore, the CM temperature with the 2-level digital modulation can be described as\cite{Gieseler2012PFC}
\begin{equation}
T_{c m}=T_0\frac{\Gamma_{0}}{\Gamma_{0}+ \delta \Gamma}=T_0\frac{\Gamma_{0}}{\Gamma_{0}+ \eta \omega_{0}/{\pi}}\text{,}
\label{T}
\end{equation}
where $T_0=298$K is the temperature of surrounding environment. In the experiment, the total damping rate $\Gamma_{t o t}=\Gamma_{0}+\delta \Gamma$ and CM temperature $T_{c m}$ can be obtained by fitting the power spectral density of the trajectory with\cite{Li2013}
\begin{equation}
S(\omega)=\frac{2 k_{B} T_{c m}}{m} \frac{\Gamma_{t o t}}{\left(\omega_{0}^{2}-\omega^{2}\right)^{2}+\omega^{2} \Gamma_{t o t}^{2}}\text{,}
\end{equation}
where $k_B$ is the Boltzmann constant.

\begin{figure}[t]
	\includegraphics[width=0.45\textwidth]{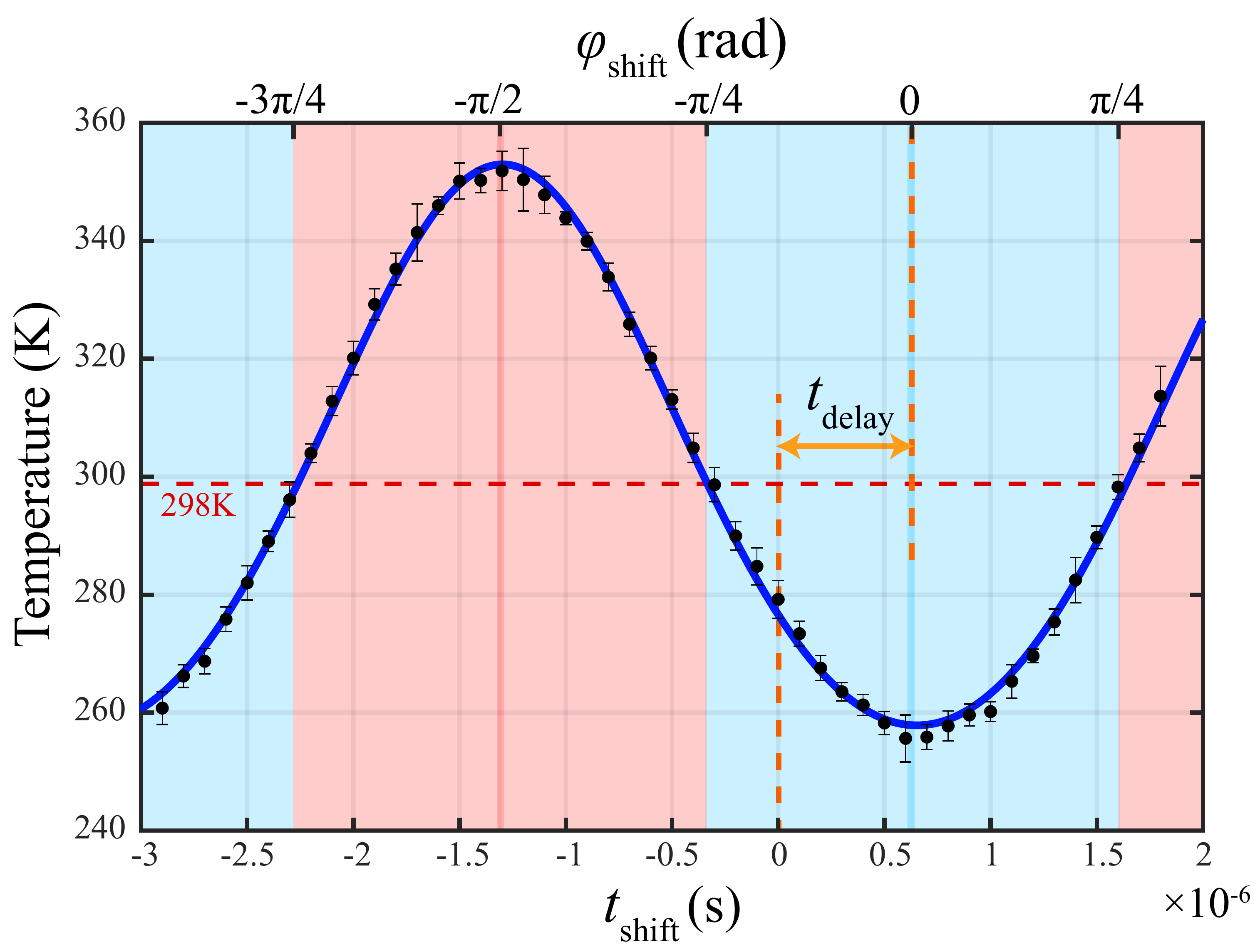}
	\caption{CM temperature changes with different shift times (phases) with the modulation depth $\eta$=0.52\% at 2 mbar. The solid line is a fit according to Eq.(\ref{TP}).}
	\label{fig:Phase}
\end{figure}

In the above consideration, only the cooling process is studied with different modulation depths. In the digital parametric feedback process, by controlling the phase shift between the motion and the modulation signal, we are able to fully control the motions of the levitated particle, in addition to the cooling. Here, we can introduce an additional phase shift $\varphi_{shift}$ to the modulation signal. The energy varying caused by modulation during one cycle of the oscillation can be rewritten as
\begin{equation}
\Delta \mathrm{E}=-m \eta \omega_{0}^{2} X_{0}^{2} \cos (2 \varphi_{shift})\text{.}
\end{equation}
Correspondingly, the CM temperature could be
\begin{equation}
T_{c m}=T \frac{\Gamma_{0}}{\Gamma_{0}+\cos (2 \varphi_{shift}) \eta \omega_{0}/{\pi}}\text{.}
\label{TP}
\end{equation}
From the above equation, when $-\pi/4<\varphi_{shift}\leq\pi/4$, the levitated particle is cooled. However, when $-3\pi/4<\varphi_{shift}\leq-\pi/4$, the levitated particle is heated. For $\varphi_{shift}=0$ and $-\pi/2$, it shows the maximal cooling and heating, respectively.


In the experiment, a 1064nm laser ($\sim$200mW) was focused by an objective (NA=0.9) in vacuum chamber to create an optical potential for particle trapping, as shown in Fig.\ref{fig:1}(b). The intensity of the trapping laser was modulated by an acousto-optic modulator (AOM). We used the zero-order light from the AOM to maximize the utilization of laser energy. A silica nano-sphere (nominal radius 82.5nm), which was sent by a nebulizer, was trapped near the focus. The forward scattering light of the trapping laser was collected by an aspheric lens (NA=0.546) and sent to three sets of balanced photodetectors (homemade, 3MHz bandwidth) to measure the positions of the trapping nano-particle along three motional degrees (set as X, Y, Z axis) of freedom\cite{Yu20193D}.

\begin{figure}[b]
	\includegraphics[width=0.45\textwidth]{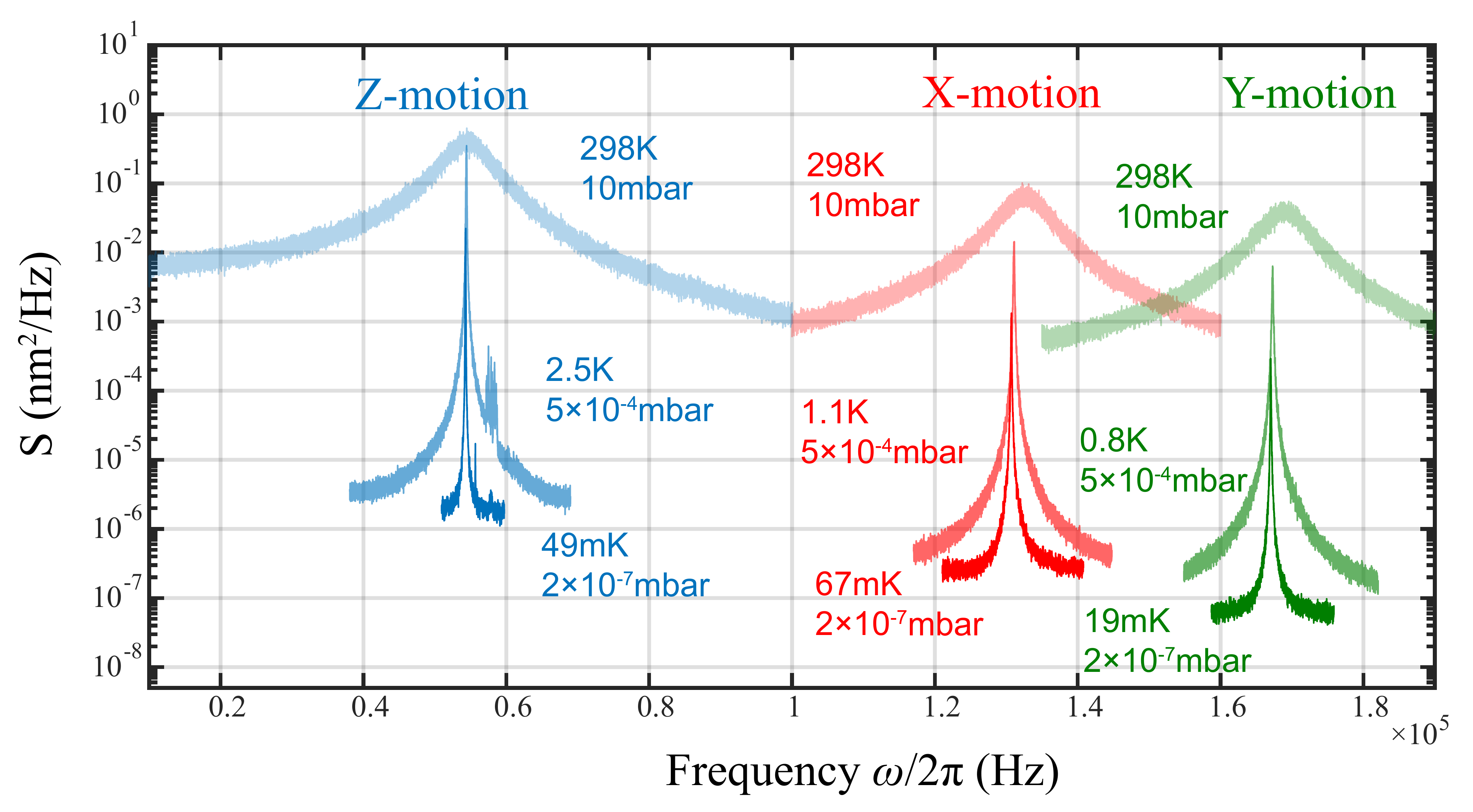}
	\caption{Power spectral densities of the feedback cooled X, Y, Z axes' motions under different pressures ($10$ mbar, $5\times10^{-4}$ mbar, $2\times10^{-7}$ mbar). The modulation depth is $\eta=0.5\%$. }
	\label{fig:PSD}
\end{figure}

The position signals were recorded by a digitizer on computer and simultaneously sent to an FPGA board to generate the feedback modulation signal. On the FPGA board, the digitized position signals were sent to Kalman filters first to reduce noise\cite{setter2018kalman,liao2018kalman}. The gain of the Kalman filter was fixed to increase processing speed. Then the filtered position signals were sent to a delay module to compensate the phase error that caused by the phase delay time $t_{delay}$ of the feedback loop. In the experimental setup, $t_{delay}=650$ ns, which was composed of $200$ ns from AOM, $160$ ns from ADC, $120$ ns from FPGA, $20$ ns from DAC, and $150$ ns from detector.

Assuming the trajectory of the oscillating particle in one axis (for example X-axis) in the feedback loop is
\begin{equation}
x(t)=A_{0} \sin \left[\omega_{x}\left(t-t_{delay}\right)+\varphi\right]\text{,}
\label{position}
\end{equation}
we can set the modulated laser intensity as
\begin{equation}
I(t)=I_{0}\left(1+0.5  \eta \times \operatorname{sign}\left[\sin \left(\omega_{x} t+\varphi\right) \cos \left(\omega_{x} t+\varphi\right)\right]\right)\text{.}
\end{equation}
In order to obtain the above modulation signal, extra delays ${\pi}/{\omega_{x}}-t_{delay}$ and ${\pi}/{(2\omega_{x})}-t_{d e l a y}$ were applied to the position signal in Eq.(\ref{position}) by the FPGA to get $\sin \left(\omega_{x} t+\varphi\right)$ and $\cos \left(\omega_{x} t+\varphi\right)$ functions, respectively.

Moreover, to verify the effect of modulation signal phase shift on the parametric feedback cooling, we can add shift time, $t_{\text {shift}}$, to the modulation signal, which can also be realized by the FPGA. Correspondingly, the modulation phase is $\varphi_{shift}=\omega_{x} (t_{\text {shift}}-t_{\text {delay}})$ at X axis.
As shown in Fig. \ref{fig:Phase}, the lowest temperature reached when $t_{\text {shift}}=t_{\text {delay}}$ in the cooling process, while the highest temperature reached when  $t_{\text {shift}}=t_{\text {delay}}-\pi /(2 \omega_{x})$ in the heating process. The solid curve in Fig. \ref{fig:Phase} is a fit of Eq.(\ref{TP}) with $T_0=298$ K, $\Gamma_{0}/2\pi=1377$ Hz, $\omega_{x}=8.054 \times 10^{5}$ rad/s, and the modulation depth $\eta=0.52\%$.



\begin{figure}[t]
	\includegraphics[width=0.45\textwidth]{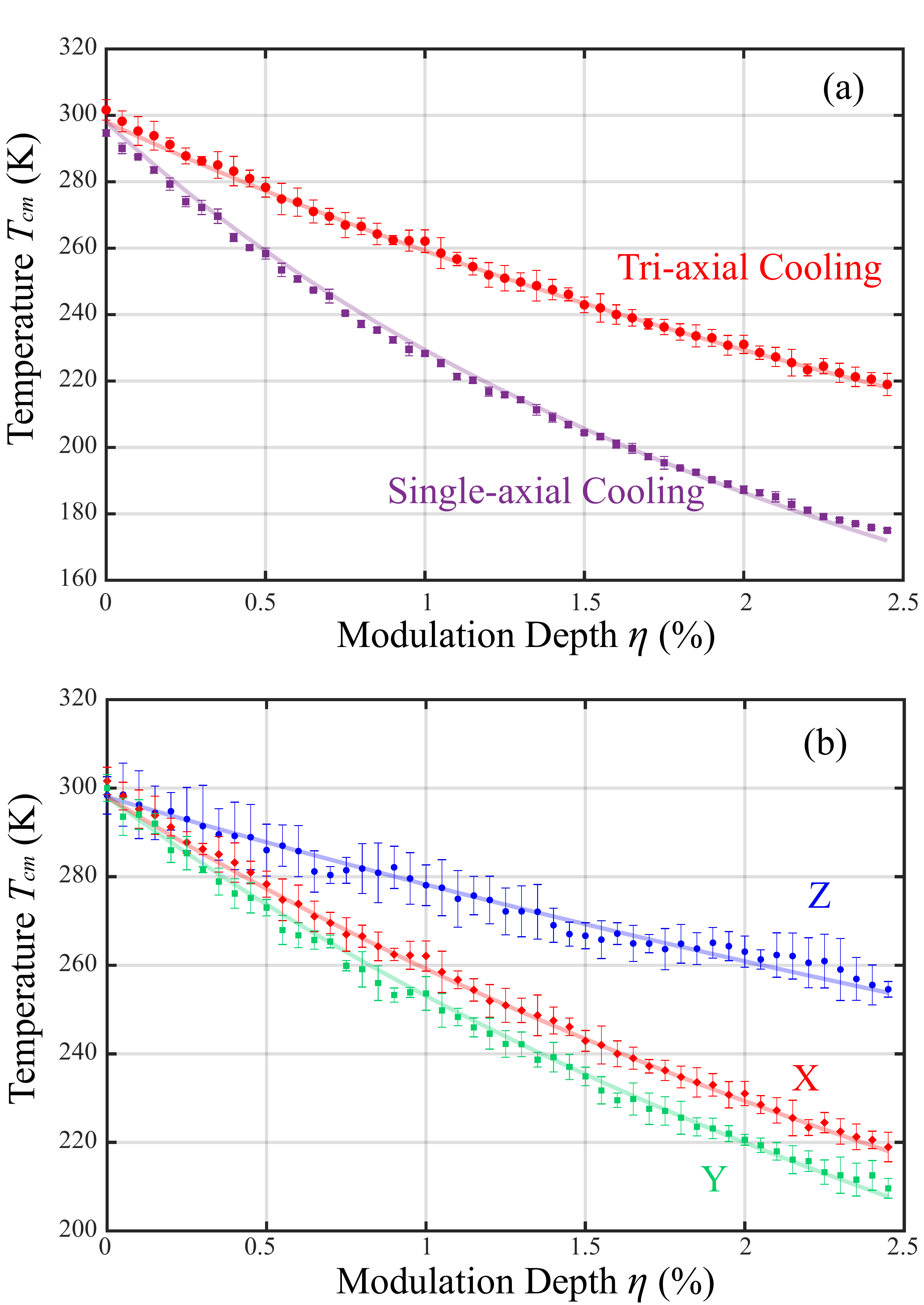}
	\caption{Relationship between CM temperature and modulation depths at $2$ mbar. (a) Comparison between single-axial cooling and tri-axial cooling. The solid lines are fits according to Eq.(\ref{T}) and Eq.(\ref{T3D}), respectively. (b) CM temperatures of X, Y, Z axes' motions under tri-axial cooling. The solid lines are fits to  Eq.(\ref{T3D}). }
	\label{fig:3}
\end{figure}

In order to cool all the three axes' motions, we chose a majority rule based on the single-axial cooling.
Because the frequencies and the phases of the three motional degrees of freedom are different, there will be three sets of incoherent 2-level digital modulation signals. To merge the three incoherent signals into a single 2-level modulation signal, we followed the majority rule that the output laser intensity level will be the same as the majority of high or low levels among the three-axis modulation requirements. However, such a strategy will reduce the efficiency of the feedback control. In the cooling of the CM temperature of the levitated particle, for each motion degree of freedom, there will be $75\%$ possibility to be cooled and $25\%$ to be heated. The overall efficiency is $50\%$ compared with the single-axis modulation. In this case, the CM temperature of each motional degrees of freedom with the tri-axial modulations should be rewritten as
\begin{equation}
T_{c m}=T \frac{\Gamma_{0}}{\Gamma_{0}+0.5 \eta \omega_{0}/{\pi}}\text{.}
\label{T3D}
\end{equation}

Figure \ref{fig:PSD} shows the power spectral densities of the cooled X, Y, Z axes' motions, demonstrating the results of the tri-axial digital feedback cooling under different pressures. By increasing the modulation depth, we can reduce the CM temperature, as shown in Fig.\ref{fig:3}. From Fig.\ref{fig:3}(a), the efficiency of single-axial cooling is higher than the tri-axial cooling, since it cools the levitated particles to a lower temperature with same modulation depth.
\begin{figure}[b]
	\includegraphics[width=0.45\textwidth]{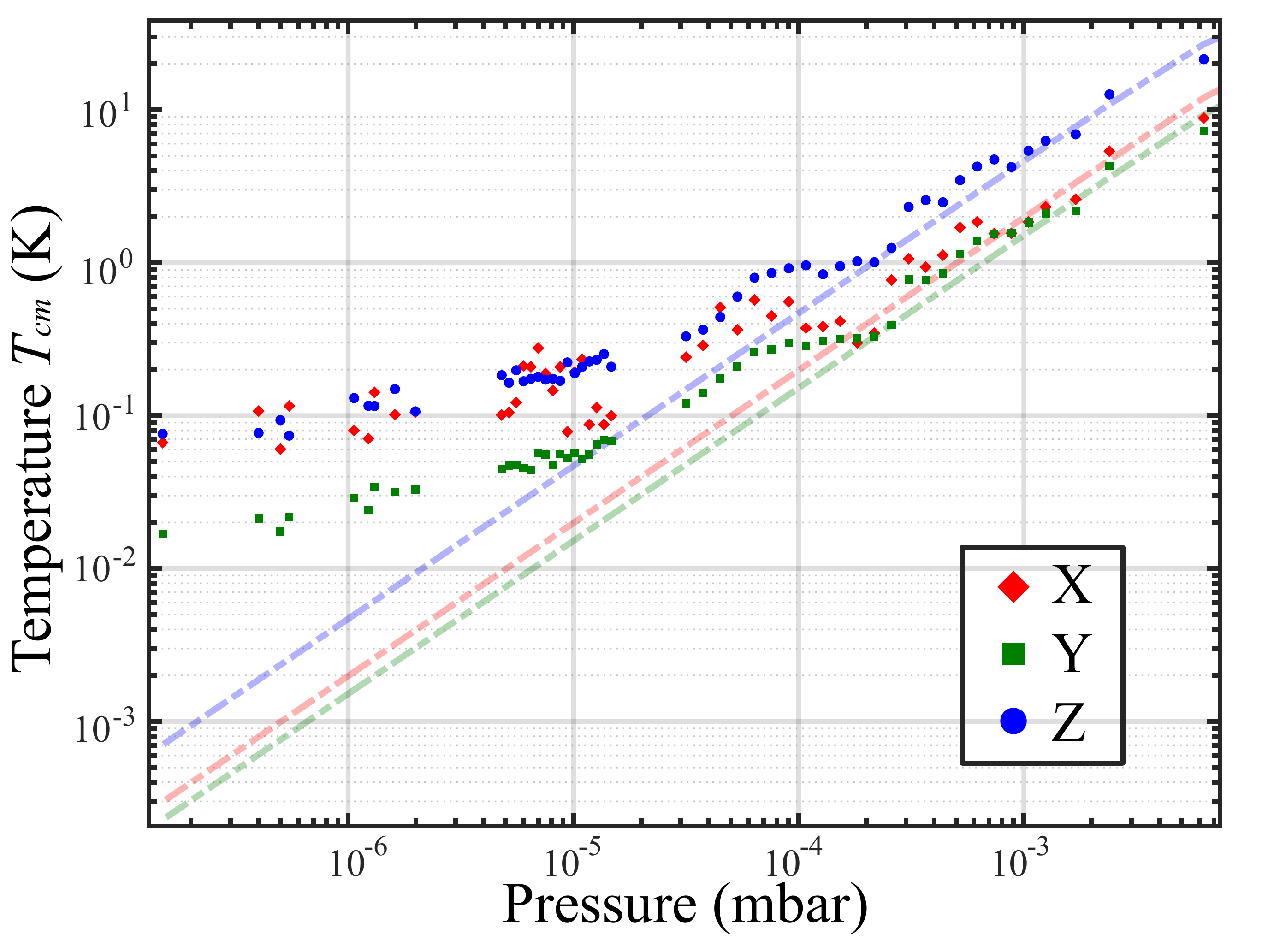}
	\caption{CM temperatures of X, Y, Z axes' motions during the decrease of air pressure. The dashed lines are the plotting of Eq.(\ref{T3D}) with air damping rate $\Gamma_{0}$ proportional to air pressure in high vacuum. The modulation depth is $\eta$=0.5\%.}
	\label{fig:4}
\end{figure}
To investigate the performance of the 2-level digital parametric feedback cooling in high vacuum, we continue to exhaust the vacuum chamber and record the varying of CM temperature and total damping rate during the decrease of air pressure. As shown in Fig. \ref{fig:4}, the temperature drops with the decrease of air pressure. However, the temperature is significant higher than the dashed line from Eq.(\ref{T3D}) when the pressure is lower than $10^{-5}$ mbar. Under high vacuum, the air damping $\Gamma_{0}$ is significant smaller than the feedback modulation damping $\delta \Gamma$. It means that the total damping $\Gamma_{tot} \approx \delta \Gamma$ should remain unchanged in high vacuum in the experiment. However, the total damping drops when the pressure is decreasing. The reason is that to make the Kalman filter feasible for a large range of pressure, the pass bandwidth is set to around 10kHz. It means that when the vibration amplitude is very small, noise can interfere with real signal, causing a decline in $\delta \Gamma$ and modulation efficiency.

\begin{figure}[t]
	\includegraphics[width=0.45\textwidth]{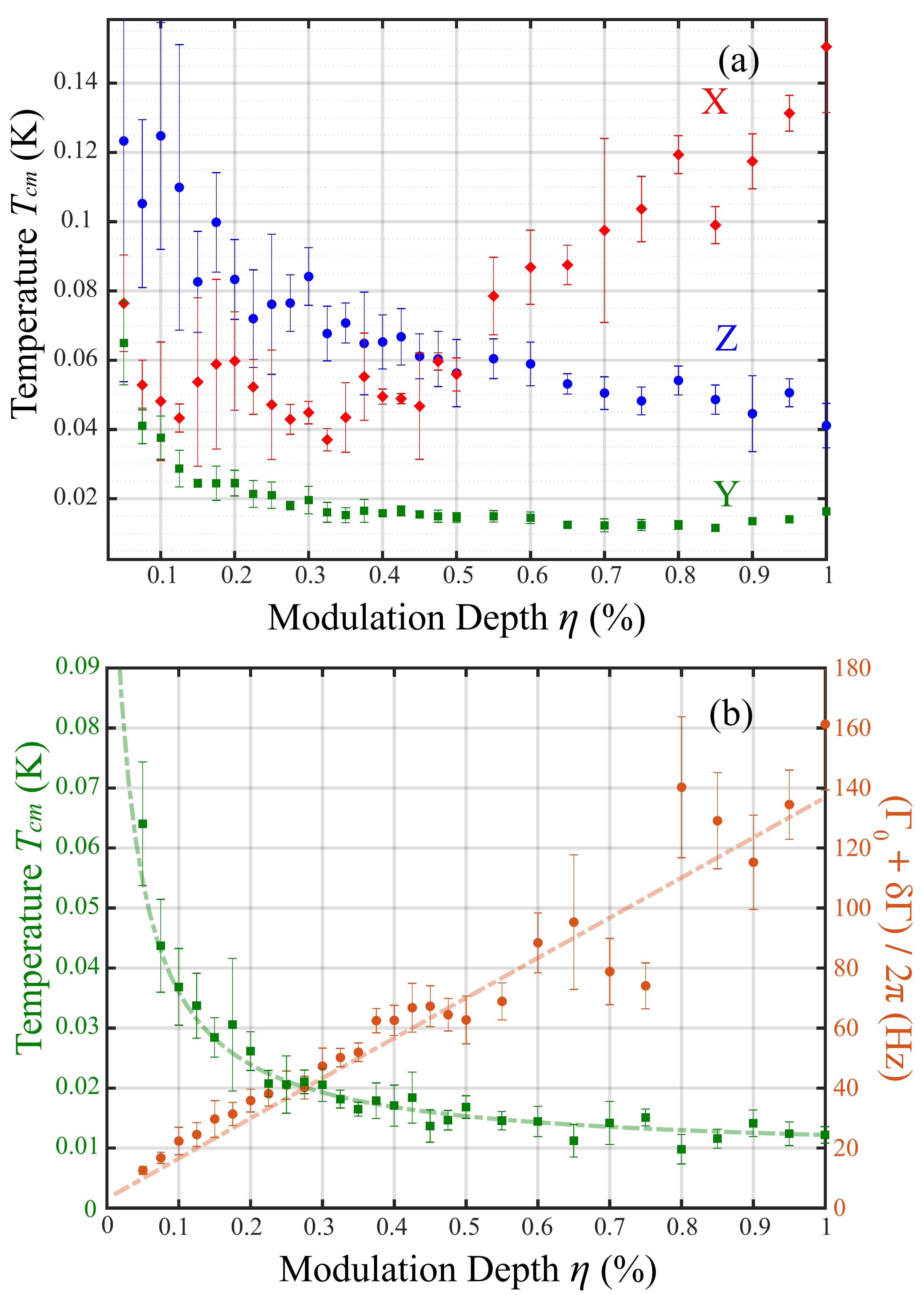}
	\caption{Cooling result versus modulation depth $\eta$ at $2 \times 10^{-7}$ mbar. (a) CM temperature $T_{cm}$ for the motions in the X, Y, Z axes versus modulation depth $\eta$. (b) CM temperature $T_{cm}$ and total damping rate $\Gamma_{t o t}$ for the motion in the Y-axis versus modulation depth $\eta$. The orange dashed line is a linear fit. The green dashed line is a fit to Eq.(\ref{T3DS}) with $\Gamma_{\mathrm{heat}}/{2\pi}=0.0016+0.4\eta$ Hz, $\beta=0.49$.}
	\label{fig:5}
\end{figure}

However, the decrease of $\delta \Gamma$ is not the only reason that the actual cooling temperature deviates much from the theoretical value. In order to explore possible reasons, we recorded the CM temperature and total damping rate with different modulation depths at $2 \times 10^{-7}$ mbar. Here, we focused on the result of Y-axis motion in Fig. \ref{fig:5}(b). The total damping rate rises as the modulation depth increases. And the temperature of the levitated particles drops as the modulation depth increases at small depth regime. But when the modulation depth keeps on increasing, the decreasing rate of temperature reduces and the temperature reaches the lowest limit of approximately $10$ mK. This indicates that modulation introduces an additional heating effect. Moreover, such an effect to heat the CM temperature is proportional to the modulation depth. This can be well observed in Fig. \ref{fig:5}(a), where the temperature of X-axis motion increases with large modulation depth. This is because that the direction of sound wave propagation in AOM is parallel to X-axis, which introduces a fluctuation of laser intensity distribution along X-axis during the modulation. Therefore, it causes a much more significant heating process compared to Y and Z-axis.

Based on previous investigation and the experiment result, we can modify Eq.(\ref{T3D}) to a high vacuum version, which is
\begin{equation}
T_{\mathrm{cm}}=T_{0} \frac{\Gamma_{\mathrm{heat}}}{\Gamma_{\mathrm{heat}}+0.5 \beta \eta \omega_{0}/{\pi}} \text{,}
\label{T3DS}
\end{equation}
where $\Gamma_{\text { heat }}=\Gamma_{0}+\Gamma_{\text { recoil }}+\Gamma_{\text { sys }}+\alpha \eta$. Here, $\Gamma_{\text { recoil }}$ is the recoil heating\cite{jain2016recoil}. $\Gamma_{\mathrm{sys}}$ is the heating caused by the instability of experiment system. $\alpha$ is the coefficient that scales the heating from modulation. $\beta$ represents the decrease of modulation efficiency under low temperature. Such an equation can well describe the temperature of the levitated particles at low pressure, as shown in Fig.\ref{fig:5} (b).

In conclusion, we have introduced a 2-level digital parametric feedback control scheme for optical levitation. It can much reduce the complexity and difficulty in the construction of feedback control. We have analyzed the effect of modulation parameters and verified it with experiment.
The coolest temperature approached $10$ mK. 
This cooling limit can be further improved with a tunable Kalman filter and laser mode filter such as single-mode fiber coupling after AOM in the future. Besides cooling and heating, such a digital feedback modulation with high programmability can be further applied to fully control the motions of levitated particle.

\section*{Acknowledgment}
This work is supported by the Science Challenge Project (No. TZ2018003), the National Natural Science
Foundation of China (Nos. 91536219, 61522508, and 91850102), the Anhui Initiative in Quantum Information
Technologies (No. AHY130000).

\providecommand{\noopsort}[1]{}\providecommand{\singleletter}[1]{#1}%
%

%
\end{document}